\documentclass[toc]{PoS}

\title{Quantum properties and generalised T-duality of the Yang-Baxter Wess-Zumino model}

\ShortTitle{Quantum and duality aspects of the Yang-Baxter Wess-Zumino model}

\usepackage{changepage,scrextend}

\usepackage{amsmath,amssymb,euscript,array,mathrsfs,appendix,ctable,marvosym,bbm,amsthm,fontenc, times,mathptmx}

\author{\speaker{Saskia Demulder}\\
        Theoretische Natuurkunde, Vrije Universiteit Brussel \& The
        International Solvay Institutes,\\ B-1050 Brussels, Belgium,\\\noindent
         Department of Physics, Swansea University, Swansea, SA2 8PP, U.K. \\
        E-mail: \email{Saskia.Demulder@vub.be}}

\abstract{In this short proceedings we discuss some of the results obtained in \cite{Demulder:2017zhz}. Integrable deformations enlarge the landscape and understanding of integrable models and its algebraic structures like quantum groups. In this short proceedings, we will review the one-loop renormalisation group analysis of an integrable deformation known as the Yang-Baxter Wess-Zumino model. This classically integrable model shows a striking stability under one-loop renormalisation. In addition, we show how Poisson-Lie T-duality, a generalisation of T-duality that is closely intertwined with integrable deformations, is particularly simple and elegant for the Yang-Baxter Wess-Zumino model.}

\FullConference{Corfu Summer Institute 2018 "School and Workshops on Elementary Particle Physics and Gravity"\\
		(CORFU2018)\\
		31 August - 28 September, 2018\\
		Corfu, Greece}
		
\begin{document}

\maketitle

\newpage

\section{Introduction}

Integrable systems are infamously known to be quite sparse in classical mechanics as well as field theory. Indeed, typically one can expect a system to be integrable when it displays a very large symmetry group. One could wonder whether known integrable models can be deformed and thus breaking\footnote{That even after deformation the system remain integrable is remarkable since . However we will see below that the symmetry groups are not so much broken as they are enhanced.} some symmetry in the process, but preserving integrability. This is most notably exemplified by integrable deformations of certain well-known integrable models. Two foremost examples of such integrable deformations are the $\eta$- and $\lambda$-deformation \cite{Klimcik:2002zj,Klimcik:2008eq,Sfetsos:2013wia}. In particular, as we will be concerned with a related deformation, the $\eta$-model is an integrable deformation of the Principal Chiral model (PCM).  The PCM and the $\eta$- and $\lambda$-models are (classically) integrable as their equations of motion can be recast into a Lax pair. It is this Lax pair that accounts for the existence of an infinite tower of conserved charges required for an integrable field theory.

The motivation to consider integrable deformations is not solely restricted to the exploration of new integrable models with \textit{less} Lie group symmetry. Indeed it has proven to be critical in the providing explicit examples of generalised T-dualities \cite{Sfetsos:2015nya,Klimcik:2015gba} and understanding the role of quantum groups in integrable systems \cite{Delduc:2013fga,Delduc:2017brb}. Both these aspects will be discussed in this talk for the particular case of an integrable deformation called the \textit{Yang-Baxter Wess-Zumino (YB-WZ) model} \cite{Delduc:2017fib}. The YB-WZ model is integrable as its equation of motion can be put into a Lax form.
The notion of integrability captured by the Lax criterion however is only valid at the classical level. Indeed, it remains an outstanding challenge to show, given a classically integrable model, whether the property of integrability survives at the quantum level. A first non-trivial check is to investigate the stability at one-loop. It is one of the aims of this talk to outline this very feature for the YB-WZ model reported in \cite{Demulder:2017zhz}.

In the first section we will introduce the Yang-Baxter Wess-Zumino model and its integrability properties. We discuss under what condition this model can be consistently renormalised at one loop and show a surprising (and suggestive?) interplay between this very quantum condition and the condition of classical integrability. The second section will discuss some relations between integrable deformations and certain quantum groups. A surprising property of the YB-WZ model is to feature a very simple Poisson-Lie duality. This will be discussed in the third section.

%%%%%%%%%%%%%%%%%%%%%%%%%%%%%%%%%%%
%%%%%%%YB-WZ and integrability%%%%%%%
%%%%%%%%%%%%%%%%%%%%%%%%%%%%%%%%%%%
\section{YB-WZ model: integrability and one-loop renormalisation group flow}

The Principal Chiral Model is a canonical example of an integrable system as it admits a Lax pair representation. Consider a Lie group $G$ and its Lie algebra $\frak g$. The action of the PCM is defined in terms of a map $g:\Sigma \rightarrow G$ from the worldsheet to the target space manifold $G$,
 \begin{eqnarray}
\label{eq:act_PCM}
  &&S_\mathrm{PCM}= -\frac{1}{2\pi}\int d \sigma d\tau    \langle g^{-1} \partial_+g , g^{-1} \partial_- g \rangle\,.
\end{eqnarray} 
One of the crucial properties of this action is that it admits a large group of symmetry given by the left and right multiplication by a group element, we will denote this global  symmetry by $G_L\times G_R$. Note that this model is not conformal without an additional Wess-Zumino term, leading to a WZW-model \cite{Witten:1983ar}. Not withstanding its lack of conformality the 2d PCM was shown in \cite{Zakharov} to allow for a Lax formulation and is thus exactly solvable.

\subsection{The YB-WZ model as an integrable model}

The possibility of deforming the action of the PCM \eqref{eq:act_PCM} whilst preserving integrability is capacitated by a special linear operator acting on the Lie algebra $\frak g$. For any semi-simple Lie algebra there exists a canonical skew symmetric endomorphism $\mathcal R: \frak g \rightarrow \frak g$ obeying the following algebraic relation,
 \begin{align} \label{eq:mcybe}
[{\cal R} x,  {\cal R} y]  - {\cal R} \left( [x, {\cal R} y ] + [ {\cal R}x, y] \right)  = [x, y] \quad  \forall x, y \in \frak{g}  \ , 
 \end{align} 
 known as the \textit{modified classical Yang-Baxter equation} (mCYBE for short) and $\cal R$ is called a (classical) R-matrix. In the following we will enforce the additional requirement that $\mathcal R^3 =-\cal R$. A geometrical interpretation of the presence of the $R$-matrix controlling the deformation will be delayed until the section about Poisson-Lie T-duality. 
  
 With an R-matrix solving \eqref{eq:mcybe} at hand, let us introduce the following deformation of the PCM known as the Yang-Baxter Wess-Zumino model \cite{Delduc:2017fib},
 \begin{eqnarray}
\label{eq:act}
  &&S= -\frac{1}{2\pi}\int d \sigma d\tau    \langle g^{-1} \partial_+g , \left(\alpha \mathbbm{1} + \beta {\cal R}  + \gamma  {\cal R}^2 \right)g^{-1} \partial_- g \rangle  \nonumber\\
&&\qquad\qquad + \frac{  k}{24\pi } \,      \int_{M_3}     \langle   \bar g^{-1} d\bar g, [\bar g^{-1} d\bar g,\bar g^{-1} d\bar g]  \rangle\,.
\end{eqnarray} 
The last term is the well-known Wess-Zumino term and as usual $k$ has to be an integer for the action to make sense in the path integral. 

Exploiting the modified Yang-Baxter equation \eqref{eq:mcybe} and the relation $\mathcal{R}^3=-\mathcal R$ satisfied by the R-matrix $\cal R$, the deformed model \eqref{eq:mcybe} can be shown to admit a Lax pair representation, provided that the values of $(\alpha, \beta, \gamma)$ are constrained to a two-dimensional submanifold of the parameter space \cite{Delduc:2014uaa},
   \begin{align}
   \label{eq:intlocus}
   \beta^2 = \frac{\gamma}{\alpha} \left( \alpha^2 - \alpha \gamma - k^2 \right) \, .
   \end{align}
   In the following this relation will be referred to as the \textit{integrable locus}.

%%%%%%%%%%%%%%%%%%%%%%%%%%%%
%%%%%%%%%%%%%%%%%%%%%%%%%%%
   \subsection{One-loop renormalisation group flow}
   In \cite{Demulder:2017zhz} the beta-function for the couplings $(\alpha, \beta, \gamma)$ appearing in the action \eqref{eq:mcybe} were computed in all generality\footnote{In \cite{Kawaguchi:2011mz} the RG behaviour has been investigated for $G=SU(2)$. However this special case thus not capture the full deformation, since for $G=SU(2)$ one of the contributions of one of the coupling terms in the action is a total derivative in the Lagrangian.}, that is without imposing integrability by constraining the couplings to stay on the integrable locus \eqref{eq:intlocus}.

  Demanding that the beta-function are consistent with the non-linear $\sigma$-model ansatz, so that the model is one-loop renormalizable, one can one can distinguish two different scenarios:
  \vspace{-10pt}                                                                                                                                                                                                                                                                                                  
\begin{enumerate}
\item Restrict the parameters to the integrable locus, keeping the group structure general, or;
\item Restrict to simply laced groups, but not restricting the parameter space.
\end{enumerate}
  \vspace{-10pt} 
  In this talk, whose objective is to outline the integrable aspects of the integrable model \eqref{eq:mcybe}, we will restrict our intention to the first scenario. Let us pause and appreciate the surprising conclusions that this former scenario seems to suggest. Indeed, whilst the integrable locus enforces \textit{classical} integrability, the condition resurges when looking at the renormalization, a purely quantum feature of the theory. Keeping in mind that showing quantum integrability is quite involved, this unexpected relation can be seen as a first hint to the quantum integrability of this integrable deformation.

\begin{figure}
\centering
\includegraphics[scale=0.83]{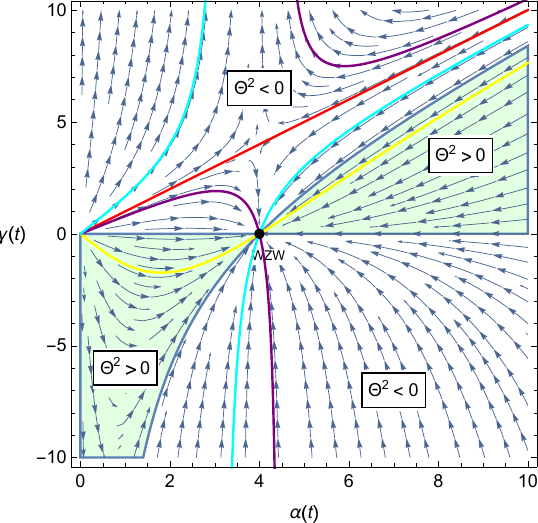}
\caption{{RG flow for $G=SU(3)$ YB-WZ model at level $k=4$ restricted to the integrable locus.}
 }\label{fig:RGSU3}
\end{figure}

 \noindent 
  Choosing to express the parameter $\beta$ in terms of the latter two,
  \begin{eqnarray}
\beta= \pm \sqrt{\frac{\gamma }{\alpha }}\, \sqrt{ \alpha ^2- \alpha \gamma -k^2}\, ,
\end{eqnarray} 
the beta-function for the remaining parameters read,
 \begin{eqnarray}
\frac{d \alpha}{dt} &=& - \frac{c_G}{2}\, \frac{k^2- \alpha ^2}{( \alpha- \gamma )^2}\,,\nonumber\\
\frac{d \gamma }{dt} &=&-\frac{c_G}{2}\, \frac{ \gamma }{\alpha ( \alpha- \gamma )^2}\, \Big(2 \alpha \gamma -3 \alpha ^2+k^2\Big)\,,\label{RGinte}
\end{eqnarray} 
here $c_G$ is the Coxeter number of the Lie group $G$.
Using the expression of the flow of $\alpha$ and $\gamma$ one can check that the flow at one-loop consistently preserves the integrable locus \eqref{eq:intlocus}. Besides the integer level $k$, the flow admits another RG invariant\footnote{In addition we will require for the action to be real and to obtain the correct sign for the kinetic term that the parameters take value in the respective domains: ($\alpha \in [|k|,\infty[ \text{ and } \gamma \in [0,(\alpha^2-k^2)/\alpha^2]$) or $(\alpha \in [0,|k|] \text{ and } \gamma \in [-(k^2-\alpha^2)/\alpha^2,0] $. Note that when $\alpha=|k|$, the integrable locus collapses into a point: the WZW action.},
\begin{align}
	\label{eq:RGinv}
\Theta^2 = \frac{\alpha  \left(\alpha ^2-\alpha  \gamma -k^2\right)}{\gamma } \ ,
\end{align}
that can used to parametrized the trajectories in the RG diagram. Fig. \ref{fig:RGSU3} shows a representative RG-flow. The physically allowed regions is the subset of the parameter space indicated in green, where $\beta$, or equivalently $\Theta$, is real valued. Although at first sight the physical parameter space seem to be divided into two distinct regions, as will become clear in section \ref{Sec:PLTDYBWZ}, these regions are in fact related  by PL duality. The RG trajectories are labelled by the RG invariant $\Theta$. A representative physical trajectory is indicated in yellow. The RG flow is controlled by the WZW fixed point in the IR.

%%%%%%%%%%%%%%%%%%%%%%%%%%%%%%%%%%%%%%%%
%%%YB-WZ and quantum group structure%%%%
%%%%%%%%%%%%%%%%%%%%%%%%%%%%%%%%%%%%%%%%
\section{YB-WZ and quantum group structure}
Integrable models generically feature an infinite tower of \textit{non-local} charges \cite{Luscher,Bernard:1990jw}. Unlike the more familiar local charges, non-local charges are no longer additive. The term `non-local' refers that local information is no sufficient to reconstruct them as their charge density is itself an integral. Generically, non-local charges will thus no longer be additive. The lack of additivity leads to a Hopf algebraic objects and the concept of quantum groups originally described by Drinfel'd \cite{Drinfeld:QG}.

\subsection{The quantum group structure of integrable deformations}
 The Principal Chiral Model is a well-studied example of a system where non-local charges appear. The PCM admits an infinite tower of non-local charges under the form of a so called Yangian, a particular instance of a quantum group\footnote{Despite what its name might suggest, a quantum group is nor a group (but an algebra) nor a quantum object. Indeed quantum group is the algebraic structure present at the level of the \textit{classical} action. In this context `quantum' refers to a non-(co)commutative deformation of the underlying Hopf algebra.}. Concretely that means that particles do not only form irreducible representations of the left-right symmetry group $G_L\times G_R$ but should be promoted to representation of their Yangian $\mathcal Y(G)_L\times \mathcal Y(G)_R$. 
 
A rigorous discussion of the Yangian structure is beyond the scope of this short note, the interested reader is referred to \cite{Loebbert:2016cdm}.

Considering the particular deformation at hand \eqref{eq:act}, it is not immediately obvious whether the Yangian symmetry of the PCM survives the integrable deformation and if not what form it would take on. The quantum group structure of the $\eta$-deformed model with WZ-term was unraveled for $G=SU(2)$ in \cite{Kawaguchi:2011mz,Kawaguchi:2013gma}. It was shown hthow the left symmetry remains unchanged, whilst the right one is now generated by charges leading to an affine quantum group\footnote{Or rather some variation thereof. Indeed the deformations of the right symmetry group generated by the addition of the WZ term is not readily identifiable with a particular known algebraic object but can be best understood as the quantum deformation of the universal enveloping algebra of the affine extension of $su(2)$. For the sake of clarity we will omit this subtlety and restrict the discussion to the normal algebra $\frak g$ rather than its affine extension.}. 

Indeed, the deformation introduced for example in the deformation \eqref{eq:act} is not so much a breaking as it is an enhancement of the underlying charge algebra to that of a quantum group. Concretely, the deformation leaves the left Yangian $\mathcal Y(\frak g)_L$ invariant whilst turning the right Yangian $\mathcal Y(\frak g)_R$ into a so-called quantum deformation of the universal enveloping algebra, 

also known as a \textit{quantum group} $\mathcal U_q(\hat{\frak{g}}) $. In summary:
\begin{align*}
	&G_L\times G_R \quad \longrightarrow \quad \mathcal Y (g) \times \mathcal Y (g) \quad \xrightarrow{\text{deform}} \quad \mathcal Y (g) \times U_q (\hat{g})\,.\\
	&{\color{gray}\;\;\,\text{\small (PCM)}\qquad \qquad \qquad \;\;\,\text{\small (PCM)}\qquad \qquad \quad \,\,\, \;\;\,\text{\small (deformed PCM)}}
\end{align*}
The letter $q$ designates the deformation complex valued parameter, it can entirely be expressed in terms of the RG invariant $\Theta$ introduced in \eqref{eq:RGinv} and the level $k$,
\begin{align}\label{Eq:qparam}
 q = \exp\left[ \frac{ 8 \pi \Theta}{\Theta^2 + k^2}\right] \ .
\end{align}
In particular the quantum group parameter is RG-invariant. It is useful to pause to consider the simple but illustrative example of the $\mathcal U_q(sl(2))$ quantum group \cite{Chari}:

	\begin{addmargin}{1cm}
	\small
 The algebra $\frak g= sl(2)$ is generated by elements $E^\pm$ and $H$ satisfying the commutation relations
\begin{align*}
	[E^+,E^-]=H\,, \quad [H,E^\pm]=\pm 2E^\pm\,.
\end{align*}
It turns out that one can perform a one-parameter deformation of the commutation relation of\footnote{Technically we should rather first promote $\frak g$ to its universal enveloping algebra.

} $sl(2)$ whilst remaining an algebra with commutation relations
\begin{align*}
	[E^+,E^-]=\frac{e^{qH}-e^{-qH}}{e^{q}-e^{-q}}\,, \quad [H,E^\pm]=\pm 2 E^\pm\,,
\end{align*}
where $q$ is the parameter measuring the deformation away from the `classical' setting since upon taking the limit $q\rightarrow 0$ the original algebraic structure is recovered,
\begin{align*}
	\lim_{q\rightarrow 0} \,[E^+,E^-]=H\,,
\end{align*}
The algebraic object determined by these commutation relations is the quantum group  $U_q(sl(2))$.
		\end{addmargin}

%%%%%%%%%%%%%%%%%%%%%%%%%%%%%%%%%%%%%%%%
%%%%%%Poisson-Lie self-T-duality%%%%%%%%
%%%%%%%%%%%%%%%%%%%%%%%%%%%%%%%%%%%%%%%%
\section{Poisson-Lie self-T-duality of the YB-WZ model}
In this section we will shortly sketch the generalisation of (abelian) T-duality known by Poisson-Lie (PL) T-duality and its first order formulation, the so-called $\mathcal E$-model. Poisson-Lie T-duality has proven to play an important role in certain integrable deformation by relating seemingly different deformations. A distinguished pair of Poisson-Lie dual integrable deformations are the $\eta$-model and the $\lambda$-model (modulo an analytic continuation in the coupling constants). Since the YB-WZ is a generalisation of the $\eta$-model it will come to no great surprise that this model is Poisson-Lie dualisable as well. Remarkably however the YB-WZ turns out to have a particularly simple behaviour under PL T-duality: it is self-dual\footnote{Yet another example of a PL-self-dual system is the (undeformed) WZW model \cite{Klimcik:1996hp}.}. The PL T-duality transformation boils down to a mere Buscher-like relabelling of the coupling parameters.

\subsection{A short introduction to Poisson-Lie T-duality and $\mathcal E$-models}
As any generalisation, Poisson-Lie T-duality is best approached by pointing to the flaws of the usual (non)-abelian and how it aims to resolves these issues.  Consider a two-dimensional sigma model with target space a Lie group manifold $G$,
\begin{align}
	S= \int \mathrm d z \mathrm d \bar z\;E_{ij}\partial x^i \bar \partial x ^j\,, \label{Eq_sigmamodel}
\end{align}
where $E_{ij}=G_{ij}+B_{ij}$ encodes the background geometry and $g\in G$. Abelian T-duality is possible when the background allows for abelian isometry with Killing vector $v$, characterised by the condition\footnote{Strictly speaking for the $B$-field to respect the isometry one needs $\mathcal L_vB=\mathrm \mathrm d\alpha$, for some one-form $\alpha$. By picking a well-chosen gauge one can always bring the left-hand side to vanish.},
\begin{align}\label{eq:isom}
	\mathcal L_v G=\mathcal L_v B=0\,,
\end{align}
where $\mathcal L_v$ denotes the Lie derivative in the direction $v$.
Following Noether, any isometric direction $v$ in the background, implies the existence of conserved charges $J_a$, that is, 
\begin{align*}
	\mathrm d\star J_a=0\,.
\end{align*}
One can then perform a T-duality transformation along that direction and obtain a different, but equivalent, background. Unfortunately, generalising this procedure to non-abelian isometric direction is no longer a duality. In the most dramatic cases, the background obtained after T-duality may not admit any isometry. It is then no longer possible to T-dualise back. In contrast to abelian T-duality, non-abelian T-duality can no longer be seen as a true duality transformation. Notwithstanding this minor disappointment, non-abelian T-duality has proven to be a valuable solution generating technique of SUGRA. 

The different nature of abelian and non-abelian T-duality is however somewhat upsetting. Bypassing the requirement of any isometries altogether, Poisson-Lie T-duality unifies and generalises the notion of T-duality. It was Klimcik's and Severa's insight \cite{Klimcik:1995ux}, that one can weaken the requirement for the `current' $J$ be closed to instead satisfy a \textit{non-commutative conservation law},
\begin{align}
	\mathrm d \star J_a = \frac{1}{2} \tilde f^{bc}{}_a{}\star J_b\wedge \star J_c\,.\label{eq:flat_connection}
\end{align}
The relation \eqref{eq:flat_connection} demands $J$ to be a flat connection with respect to a new algebraic structure $\tilde{\frak{g}}$ with structure constants $\tilde f_a{}^{bc}$. Sigma-models admitting the existence of such currents are called \textit{Poisson-Lie symmetric}. On the level of the background fields, the condition to be Poisson-Lie dualisable, replacing the isometry condition \eqref{eq:isom}, becomes,
\begin{align*}
	&\mathcal L_{v_a}E_{ij}=\tilde f^{bc}{}_av_b{}^kv_c{}^lE_{ik}E_{lj}\,.
\end{align*}
In particular, the flatness condition implies that we can directly integrate the current in terms of a group element $\tilde g\in \exp(\tilde{\frak{g}})=\widetilde G$: $J=\tilde g^{-1}\mathrm d \tilde g$.

Note that the non-commutative conservation law \eqref{eq:flat_connection} satisfied by PL symmetric $\sigma$-models betrays the existence of a \textit{second} Lie group $\tilde G$ obtained by exponentiating the algebra with structure constants $\tilde f$ in \eqref{eq:flat_connection}. Although we will now show it here, the very existence of the flat connection $J$ and the particular form the conservation law \eqref{eq:flat_connection} implies that any extremal surface of the initial sigma-model parametrised by a group element $g(\sigma, \tau)$ can be lifted to an element $l(\sigma, \tau)= g(\sigma, \tau) \tilde g(\sigma, \tau)$, where $\tilde g(\sigma, \tau)$ solves \eqref{eq:flat_connection}, living in a larger group $D$ called a \textit{Drinfel'd double}. A Drinfel'd double is an even dimensional Lie groups equipped with an ad-invariant innerproduct $\langle\,,\,\rangle$ and such that it admits two complementary Lagrangian subgroups $G$ and $\widetilde G$, i.e. 
$ D=G\cdot \widetilde G$. A subalgebra $\frak g$ is called Lagrangian if it is half-dimension compared to the group and isotropic, that is $\langle \frak g, \frak g\rangle$ with ad-invariant non-degenerate product on $\frak d$. 

The democratic way in which both groups $G$ and $\widetilde G$ are embedded in the Drinfeld double $D$ is however very suggestive. Indeed one could wonder if starting with a sigma-model with target space $\widetilde G$ and expect that upon satisfying the Poisson-Lie condition, that dual algebra is exactly $\frak g$. This suspicion turns to be true and is nothing but Poisson-Lie T-duality relating a certain sigma-model on the target space manifold G with another sigma-model on $\widetilde G$\footnote{Beware that we will keep to the simplest setting for PL T-duality. The Drinfel'd double is assumed to be perfect, that is every element of $D$ has a unique decomposition $l=g\tilde g$ for $g\in G$ and $\tilde g \in \widetilde G$}. We will not discuss how from the initial PL-dualisable sigma-model one obtains its PL-dual sigma-model. Instead we will explain in the next section how pairs of sigma models can be produced starting from a Drinfel'd double $D$ rather than a target space manifold $G$.

\subsection{First order formulation of PL: $\mathcal E$-models}
Whilst the previous section presented a succinct overview of what we might call the \textit{bottom-up} formulation of Poisson-Lie T-duality, it will be more convenient to work in a \textit{top-down} framework where Poisson-Lie T-duality is manifest. Instead of starting from a Poisson-Lie symmetric sigma-model, the starting point is now a Drinfel'd double $D$ together with a Hamiltonian determining a first order formulation of the dynamics with symplectic form $\alpha$  and Hamiltonian $H_\mathcal{E}$  on the Drinfel'd double $D$ \cite{Klimcik:2017ken},
\begin{align}
	S_\mathcal{E}= \int \alpha-H_\mathcal{E}\mathrm d t\,, \label{eq-EmodelAction}
\end{align}
The construction of an $\mathcal E$-model relies on a particular choice of a self-adjoint involutive operator $\mathcal E$ from which the Hamiltonian in \ref{eq-EmodelAction} is obtained. The details of the construction of $\mathcal E$-models are not important for the discussion at hand and the reader is invited to consult \cite{Klimcik:2015gba,Klimcik:2017ken}.

Remember that the defining property of the Drinfel'd double is that it can be decomposed into two complementary constituent Lagrangian subgroups $D=G\widetilde G$. It turns out that by descending into the cosets $D/G$ and $D/\widetilde G$, the first order action \eqref{eq-EmodelAction} leads to a pair of Poisson-Lie T-dual sigma-models in these two target spaces.

\begin{figure*}
\centering
\includegraphics[scale=0.4]{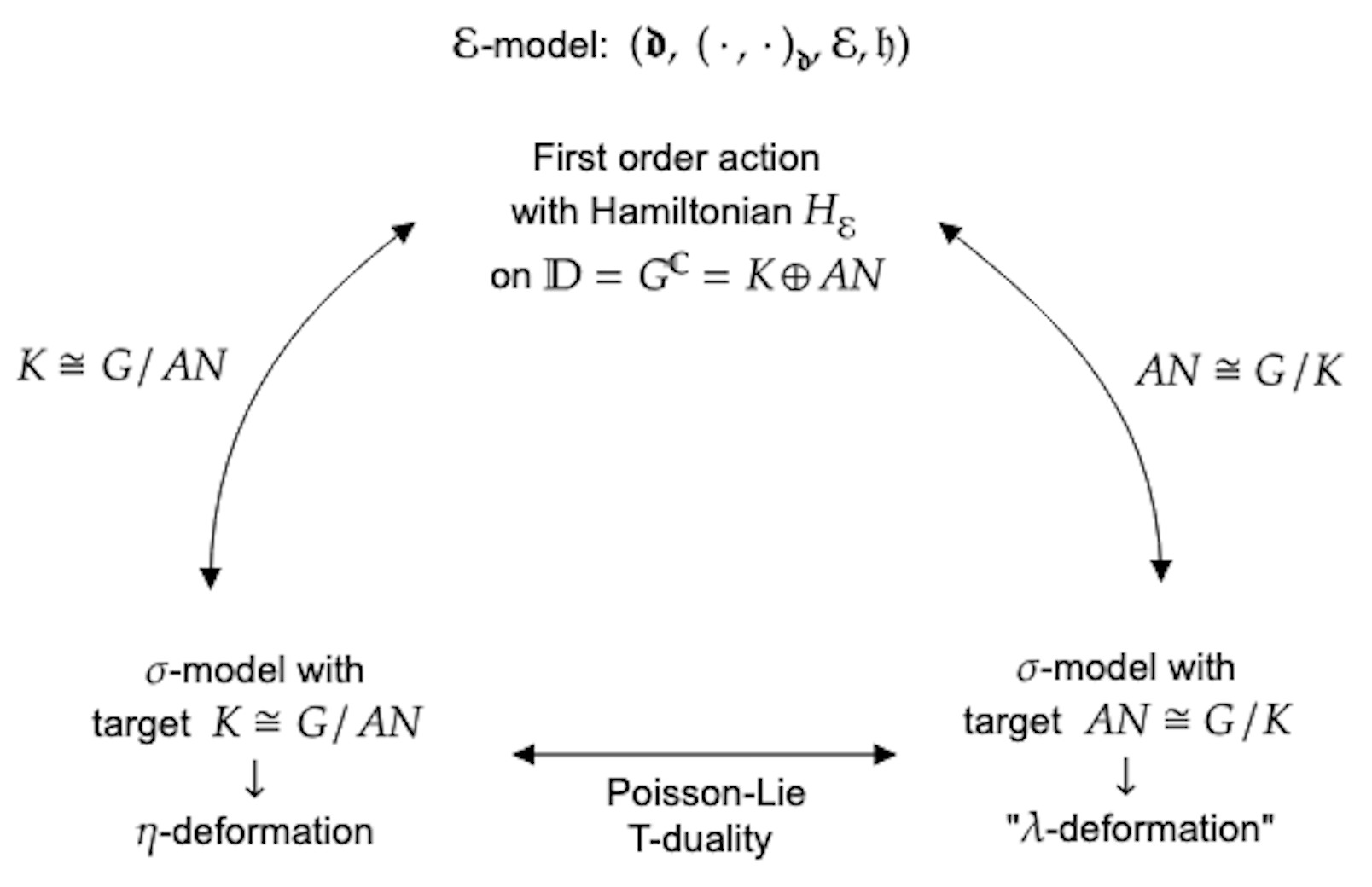}\caption{Poisson-Lie T-dual models, the $\eta$- and (analytic continuation) of the $\lambda$-model, or $\lambda^\star$-model, obtained by the same $\mathcal E$-model.}\label{Fig-PLeta-lambda}
\end{figure*}

A distinguished example is the relation between the $\eta$- and $\lambda$-deformations. In \cite{Sfetsos:2015nya,Klimcik:2015gba} it was shown that, although being seemingly very different, the $\eta$- and $\lambda$-deformation are related by Poisson-Lie T-duality together with an analytic continuation in the coupling constants. In particular, the $\eta$- and $\lambda$-deformation (modulo an analytic continuation for the latter, known as $\lambda^\star$-models \cite{Hoare:2018ebg}) can be embedded into the same $\cal E$-model. In this case the Drinfel'd double is the complexification of a compact and simple group $G$, i.e. $D=G^\mathbb{C}$. The subgroups $G$ and $\widetilde G$ can be identified by means of the Iwasawa decomposition $G^\mathbb{C}=G\cdot AN\equiv G\cdot \widetilde G$. This is diagrammatically summarised in Figure \ref{Fig-PLeta-lambda}. It turns out that the YB-WZ satisfies the Poisson-Lie condition as well and displays an especially simple behaviour under PL T-duality.

\subsection{YB-WZ model: PL-self-duality}\label{Sec:PLTDYBWZ}
To obtain \cite{Klimcik:2017ken} the YB-WZ one has, once again, to consider the Drinfel'd double $D=G^{\mathbb C}$, but this time with `twisted' Iwasawa decomposition in order to accommodate for the WZ-term in the action \eqref{eq:act},
\begin{align*}
	D=G^{\mathbb C}=G\cdot A_\rho N\equiv H_\rho \cdot \widetilde H_\rho\,,
\end{align*}
equipped with the inner product $(z_1,z_2)_\frak{d}=C\text{Im }\langle e^{i\rho}z_1,z_2\rangle$, where $C$ and $\rho$ are real numbers and elements $z\in \frak d$. %In \eqref{}, $\mathcal E$ denotes an idempotent and self-adjoint operator {\color{red} that determines}.
%
%
%
%
%\begin{center}
%\color{teal}
%$D/H_\rho$	and $D/\widetilde H_\rho$ can be identified with $G$.
%\end{center}
%
  Whilst in general applying a Poisson-Lie (or even the special case of a non-abelian) T-duality, one expects the dual background to be widely different from the initial background. Surprisingly, applying the coset reduction along the other subgroup $H_\rho$ one obtains exactly the same action modulo some Buscher-like swapping of the deformation parameters:
\begin{gather}
 \begin{aligned}\label{eq:plduality}
 \alpha &\to  \tilde{\alpha} = \frac{k^2}{\alpha}\, ,  \\
 \beta &\to  \tilde{\beta} = - \beta\, , \\
 \gamma &\to   \tilde{\gamma} = \frac{k^2 + \alpha \gamma - \alpha^2}{\alpha} = -\frac{\beta^2}{\gamma} \, . 
  \end{aligned}
\end{gather}

\begin{figure}
\centering
\label{Afb_PL_YBWZ}
\includegraphics[scale=0.38]{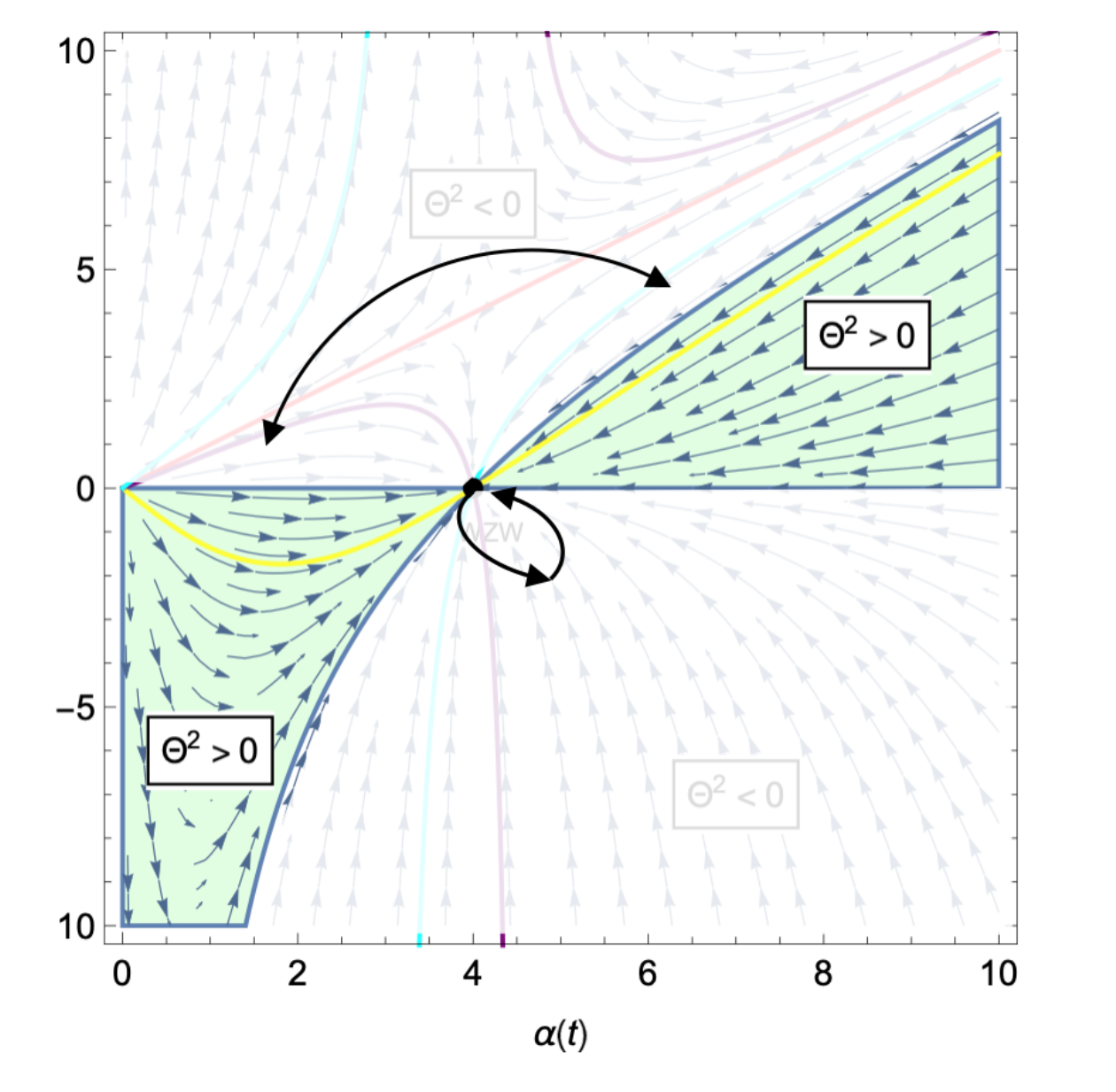}\caption{Poisson-Lie T-duality display a very intuitive action of the one-loop RG diagram. Under the effective PL-transformation, the dividing WZ-point is self-dual whilst Poisson-Lie T-duality swaps the two seemingly disconnected physical regions.
}
\end{figure}

The Poisson-Lie T-duality agreeably dovetails the one-loop renormalisation structure of the YB-WZ model:
\begin{itemize}
	\item Inspecting the expression for the integrable submanifold and the Buscher-like transformation for the PL T-dual of the YBWZ model, one can check that the integrable locus is stable under Poisson-Lie T-duality transformation. Indeed, since Poisson-Lie T-duality is a canonical transformation and therefore expected to preserve the integrable structure. Here these expectations are already met at first loop is the renormalisation group flow.
	\item Looking back at the RG-flow in Fig. \ref{fig:RGSU3}, one can see that under Poisson-Lie T-duality the apparent division between the two physical regions is deceptive. Taking  Poisson-Lie T-duality into account the regions can be identified as the duality swaps both regions. In addition we retrieve the familiar result that the WZW-model is self-dual under Poisson-Lie T-duality \cite{Klimcik:1996hp}. This is illustrated in Fig. \ref{Afb_PL_YBWZ}.
	\item  At one-loop, Poisson-Lie T-duality is compatible with the quantum group structure. Indeed the quantum group parameter \eqref{Eq:qparam} remains under invariant under the duality transformation \eqref{eq:plduality}.
\end{itemize}

%%%%%%%%%%%%%%%%%%%%%%%%%%%%%%%
%%%%%%%%%%%Discussion%%%%%%%%%%%
%%%%%%%%%%%%%%%%%%%%%%%%%%%%%%%
\section{Summary and outlook}

In this talk we have considered the one-loop behaviour of an integrable extension of the $\eta$-deformation known as the Yang-Baxter Wess-Zumino model. The one-loop RG was shown to be consistent with the integrability and quantum group properties present at the classical level. In addition we discussed how the Yang-Baxter Wess-Zumino model has proven to be a very simple avatar of Poisson-Lie T-duality. The generalised T-duality transformation preserves the form of the action and merely relabels the deformation parameters. 

It remains however to be shown whether quantum integrability remains true at all-loop. Although proving quantum integrability is infamously known to be elusive in non-ultralocal theories, of which the Yang-Baxter deformations are examples of, the approach suggested in \cite{Appadu:2017fff} might open the path towards an S-matrix formulation of the deformed models.

Yet another open avenue is to generalise the analysis of the one-loop beta-functions to the more general bi-Yang-Baxter plus Wess-Zumino term deformation \cite{Delduc:2017fib}. The additional difficulty here, is the absence of both the left- and right-actions of the initial group symmetry and might prove challenging to find an adequate basis to compute the $\beta$-functions.

%Let is end with the most ambitious and speculative 

\section*{\textit{Acknowledgments}}
I would like to thank the organisers of the Dualities and Generalized Geometries workshop of the Corfu Summer Institute 2018``School and Workshops on Elementary Particle Physics and Gravity'' for the engaging conference and the opportunity to present these results. 
I also wish to thank Sibylle Driezen, Daniel C. Thompson and Alexander Sevrin for engaging discussions and collaboration that lead to the results discussed here and I am especially grateful to S. Driezen for a careful and critical proofread of this proceedings.

This work is partially supported by the ``FWO-Vlaanderen'' through the project G006119N and by the Vrije Universiteit Brussel through the Strategic Research Program ``High-Energy Physics'' as well as through the aspirant fellowship.

%%%%%%%%%%%%%%%
%%%%%%%%%%%%%%%

\end{document}